\documentclass[RNAAS]{aastex63}

\graphicspath{{./}{figures/}}

\begin{document}

\title{No Black Holes in NGC 6397}

\correspondingauthor{Nicholas Z.\ Rui}
\email{nrui@caltech.edu}
\correspondingauthor{Newlin C.\ Weatherford}
\email{newlin@u.northwestern.edu}

\author[0000-0002-1884-3992]{Nicholas Z. Rui}
\affiliation{TAPIR, California Institute of Technology, Pasadena, CA 91125, USA}

\author[0000-0002-9660-9085]{Newlin C.\ Weatherford}
\affil{Department of Physics \& Astronomy and Center for Interdisciplinary Exploration \& Research in Astrophysics (CIERA), Northwestern University, Evanston, IL 60208, USA}

\author[0000-0002-4086-3180]{Kyle Kremer}
\altaffiliation{NSF Astronomy \& Astrophysics Postdoctoral Fellow}
\affiliation{TAPIR, California Institute of Technology, Pasadena, CA 91125, USA}
\affiliation{The Observatories of the Carnegie Institution for Science, Pasadena, CA 91101, USA}

\author[0000-0002-3680-2684]{Sourav Chatterjee}
\affil{Tata Institute of Fundamental Research, Homi Bhabha Road, Mumbai 400005, India}

\author[0000-0002-7330-027X]{Giacomo Fragione}
\affil{Department of Physics \& Astronomy and Center for Interdisciplinary Exploration \& Research in Astrophysics (CIERA), Northwestern University, Evanston, IL 60208, USA}

\author[0000-0002-7132-418X]{Frederic A.\ Rasio}
\affil{Department of Physics \& Astronomy and Center for Interdisciplinary Exploration \& Research in Astrophysics (CIERA), Northwestern University, Evanston, IL 60208, USA}

\author[0000-0003-4175-8881]{Carl L.\ Rodriguez}
\affil{McWilliams Center for Cosmology, Department of Physics, Carnegie Mellon University, Pittsburgh, PA 15213, USA}

\author[0000-0001-9582-881X]{Claire S. Ye}
\affil{Department of Physics \& Astronomy and Center for Interdisciplinary Exploration \& Research in Astrophysics (CIERA), Northwestern University, Evanston, IL 60208, USA}


\begin{abstract}
  Recently, \citet{vitral2021does} detected a central concentration of dark objects in the core-collapsed globular cluster NGC~6397, which could be interpreted as a subcluster of stellar-mass black holes. However, it is well established theoretically that any significant number of black holes in the cluster would provide strong dynamical heating and is fundamentally inconsistent with this cluster's core-collapsed profile. Claims of intermediate-mass black holes in core-collapsed clusters should similarly be treated with suspicion, for reasons that have been understood theoretically for many decades. Instead, the central dark population in NGC~6397 is exactly accounted for by a compact subsystem of white dwarfs, as we demonstrate here by inspection of a previously published model that provides a good fit to this cluster. These central subclusters of heavy white dwarfs are in fact a generic feature of core-collapsed clusters, while central black hole subclusters are present in all {\em non\/}-collapsed clusters.
\end{abstract}

\section{}
\vspace{-0.7cm}

Globular clusters (GCs) are highly dynamic systems hosting a wide range of stellar phenomena, particularly involving compact objects.
Recently, analyzing a heightened central velocity dispersion in NGC~6397, \citet{vitral2021does} detected a central dark population, which they suggested may be a subcluster of stellar-mass black holes (BHs) of total mass $10^3\,M_\odot$ within the central $6$ arcsec ($0.07\,\rm{pc}$).
However, the presence of such a BH subcluster would contradict longstanding consensus from GC modeling that \textit{any} substantial population of BHs within a cluster would provide enough dynamical heating to prevent core collapse and, instead, sustain a large, easily resolvable core \citep{Merritt2004,Mackey2007,wang2016dragon,kremer2019initial}.
For lists of Galactic GCs that {\em are expected to retain a large population of BHs} at present, see \cite{Askar2018}, \cite{Weatherford2020}, and \cite{Shishkovsky2020}.

\begin{figure}
    \centering
    \includegraphics[scale=1.2]{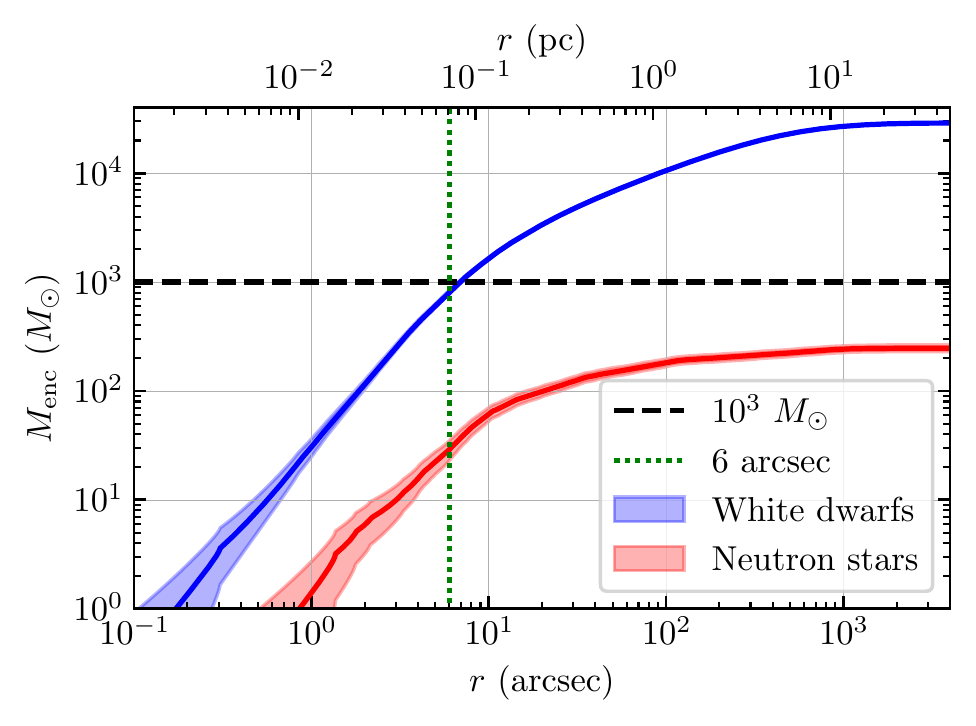}
    \caption{The enclosed mass of white dwarfs and neutron stars versus projected radius for the best-fitting NGC~6397 model \citep{Rui2021} found in the \texttt{CMC Cluster Catalog} \citep{kremer2020modeling}.
    Solid curves bounded by shaded regions indicate the expectation value and spread from all possible projections.
    In contrast to black holes, white dwarfs are a natural and obvious ``dark'' population contributing 
    $10^3\,M_\odot$ (\textit{black dashed line}) within $6$ arcsec (\textit{green dotted line}) of the cluster center, in concordance with a recent analysis of NGC~6397 data by \citet{vitral2021does}.
    Not shown here are black holes, as this model for NGC~6397 contains only a single remaining black hole, of mass $17.4$ $M_\odot$.}
    \label{fig:wd_pop}
\end{figure}

\vspace{0.2cm}
\subsection{No stellar-mass BH population in NGC 6397}

Massive star evolution produces hundreds to thousands of BHs in typical GCs, most of which (unlike neutron stars) are initially retained \citep[e.g.,][]{Kroupa2001}.
These BHs quickly mass-segregate to the cluster core, assembling a dense central BH subsystem \citep[e.g.,][]{Spitzer1969, Kulkarni1993}.
Three-body encounters within this BH-dominated core produce many dynamically-hard BH binaries \citep[e.g.,][]{Morscher2015}, which then provide energy to passing stars through scattering interactions.
This further hardens the binaries while heating the rest of the cluster \citep{BreenHeggie2013}.
BHs undergoing these binary-mediated encounters receive significant recoil kicks that 
displace them away from the core (and eventually may eject them from the cluster altogether); they further heat the cluster through dynamical friction while sinking back to the core. Overall, the dynamics of BHs acts
as a strong energy source in a process we call ``BH burning'' \citep[for a review, see][]{kremer2020tidal}. This process is well-understood and well-supported by a wide scientific consensus \citep{Merritt2004, Mackey2007,BreenHeggie2013,Peuten2016,wang2016dragon, Chatterjee2017,ArcaSedda2018,kremer2018black,kremer2019initial,AntoniniGieles2020}.

The energy generated by BH burning inflates the cluster and supports it against gravothermal contraction.
Hence, clusters that retain sizable BH populations today exhibit large core radii with flat central surface brightness profiles (well fit by King models) and reduced mass segregation in the luminous stellar populations \citep[e.g.,][]{Chatterjee2017,Weatherford2018,Weatherford2020,kremer2020modeling}.
BH burning is so powerful that star clusters exceeding a critical mass fraction in BHs may even evolve 
towards $100\%$ BH clusters, after ejecting their entire luminous stellar populations \citep{Weatherford2021,gieles2021supra}.
Importantly, a star cluster will only evolve towards a traditional ``core-collapsed'' surface brightness profile after almost all BHs have been ejected.
Since NGC~6397 is core-collapsed, it should therefore be expected to have retained very few, if any, BHs at present (the model illustrated in Fig.~\ref{fig:wd_pop} contains just one BH).

While direct probes of the BH population are sparse \citep[binary counts from X-ray/radio and/or radial-velocity measurements are instructive but typically small; e.g.,][]{Strader2012,Giesers2018}, indirect probes such as the degree of mass segregation are powerful when combined with careful modeling.
Along these lines, \citet{Weatherford2020} leverage a known anti-correlation between clusters' observable mass segregation and BH populations to place, at 95\% (67\%) confidence, restrictive upper limits of 16 (8) BHs with total mass less than $420\,M_\odot$ ($200\,M_\odot$) in NGC~6397 at present. This eliminates BHs as a plausible explanation for the central dark population in this cluster.

\vspace{0.2cm}
\subsection{No IMBHs in core-collapsed clusters}

For similar reasons, claims of intermediate-mass BHs (IMBHs) at the centers of core-collapsed GCs \citep[e.g.,][]{gerssen2002hubble,kamann2016muse,perera2017evidence} have not withstood follow-up studies and are almost certainly incorrect \citep[e.g.,][]{mcnamara2003does,murphy2011fokker,kirsten2012no,gieles2018mass,tremou2018maveric}.
Like stellar-mass BHs, an IMBH would provide a central dynamical heat source which would act to inflate the cluster core significantly \citep{shapiro1977dissolution,marchant1980star,heggie2007core}.
Clusters with IMBHs are expected to resemble standard King models except within the small radius of influence of the IMBH \citep{Baumgardt2005}.
Searches for IMBHs in core-collapsed clusters in pursuit of the theorized cusp in the surface density are therefore misguided.
On the contrary, such searches should focus on clusters which have \textit{not} undergone core collapse, where a stronger case for an IMBH might be made.

\subsection{The alternative: a massive subsystem of white dwarfs}

In Fig.~\ref{fig:wd_pop}, we illustrate a GC model that closely fits NGC~6397's surface brightness and velocity dispersion profiles.
The initial cluster contained $N=4\times10^5$ stars and was relatively compact (virial radius $r_v=1\,\mathrm{pc}$).
The present-day model agrees quite well with many observations, including numbers of cataclysmic variables, millisecond pulsars, and low-mass X-ray binaries \citep{Rui2021}.
The model belongs to a large, recently released model grid, the \texttt{CMC Cluster Catalog}, designed to broadly probe the space of realistic Milky Way GCs without any directed attempt to fit any particular cluster \citep{kremer2020modeling}.

The best-fitting model from this grid for NGC~6397 possesses only a single stellar-mass BH. However, it contains close to $10^3\,M_\odot$ in white dwarfs within the central $6$ arcsec (see Fig.~\ref{fig:wd_pop}), neatly and completely accounting for the dark population detected by \citet{vitral2021does}.
In our model, the white dwarf population is dominated by heavy white dwarfs with a mean mass $\approx1\, M_\odot$; most of these are carbon-oxygen white dwarfs ($86\%$), with a sizable minority of oxygen-neon white dwarfs ($14\%$).
Crucially, both the lack of stellar-mass BHs and the presence of a centrally concentrated white dwarf population are generic, robust features expected in all core-collapsed GCs, rather than esoteric predictions expected to apply only to NGC~6397 \citep[see, e.g.,][where this result was demonstrated generally]{kremer2020modeling}.
We will further explore the implications of these white dwarf subsystems in core-collapsed clusters in a forthcoming work (Kremer et al. in prep.).

\bibliographystyle{aasjournal}
\bibliography{noblackholes.bib}

\end{document}